\documentclass[aps,prl,superscriptaddress,twocolumn,floatfix,citeautoscript,longbibliography,hyperlinks]{revtex4-1}

\usepackage{graphicx}
\usepackage{dcolumn}
\usepackage[dvipsnames]{xcolor}
\usepackage[colorlinks,citecolor=blue,urlcolor=blue]{hyperref}
\usepackage{amssymb,amsmath,bm,bbm,mleftright,mathtools}
\DeclareMathAlphabet\mathbfcal{OMS}{cmsy}{b}{n}
\newcommand{\bra}[1]{\left\langle #1\right|}
\newcommand{\ket}[1]{\left|#1\right\rangle}

\makeatletter
\newsavebox{\@brx}
\newcommand{\llangle}[1][]{\savebox{\@brx}{\(\m@th{#1\langle}\)}%
  \mathopen{\copy\@brx\kern-0.5\wd\@brx\usebox{\@brx}}}
\newcommand{\rrangle}[1][]{\savebox{\@brx}{\(\m@th{#1\rangle}\)}%
  \mathclose{\copy\@brx\kern-0.5\wd\@brx\usebox{\@brx}}}
\makeatother

\begin{document}

\title{Quantum reservoir computing on random regular graphs}

\author{Moein N. Ivaki}
\affiliation{Quantum Technology Finland Center of Excellence, Department of Applied Physics, Aalto University,
P.O. Box 11000, FI-00076 Aalto, Finland}
\author {Achilleas Lazarides}
\affiliation{Interdisciplinary Centre for Mathematical Modelling and Department of Mathematical Sciences, Loughborough University, Loughborough, Leicestershire LE11 3TU, United Kingdom}
\author{Tapio Ala-Nissila}
\affiliation{Quantum Technology Finland Center of Excellence, Department of Applied Physics, Aalto University,
P.O. Box 11000, FI-00076 Aalto, Finland}
\affiliation{Interdisciplinary Centre for Mathematical Modelling and Department of Mathematical Sciences, Loughborough University, Loughborough, Leicestershire LE11 3TU, United Kingdom}

\begin{abstract}
Quantum reservoir computing (QRC) is a low-complexity learning paradigm that combines the inherent dynamics of input-driven many-body quantum systems with classical learning techniques for nonlinear temporal data processing. Optimizing the QRC process and computing device is complex task due to the dependence of many-body quantum systems to various factors. To explore this, we introduce a strongly interacting spin model on random regular graphs as the quantum component and investigate the interplay between static disorder, interactions, and graph connectivity, revealing their critical impact on quantum memory capacity and learnability accuracy. We tackle linear quantum and nonlinear classical tasks, and identify optimal learning and memory regimes through studying information localization, dynamical quantum correlations, and the many-body structure of the disordered Hamiltonian. In particular, we uncover the role of previously overlooked network connectivity and demonstrate how the presence of quantum correlations can significantly enhance the learning performance. Our findings thus provide guidelines for the optimal design of disordered analog quantum learning platforms.

\end{abstract}
\maketitle
{{\it{Introduction.}}} 
Quantum machine learning leverages principles of quantum computing to enhance and accelerate traditional machine learning algorithms, offering potential breakthroughs in data processing and complex problem solving beyond classical capabilities~\cite{biamonte2017quantum,RevModPhys.91.045002}. The high-dimensional space of quantum states and the rich dynamics of quantum channels offer possibilities for designing low-complexity and high-performing platforms for quantum information processing~\cite{daley2022practical, PRXQuantum.3.030101}. Quantum reservoir computing (QRC) has recently emerged as a promising approach toward temporal data processing without the need for often inefficient and inaccurate gradient optimization~\cite{fujii2017harnessing, ghosh2019quantum, markovic2020quantum, bravo2022quantum, senanian2023microwave, hu2023overcoming,mujal2021opportunities,yasuda2023quantum, kobayashi2024feedback,nakajima2020physical, ahmed2024prediction,fujii2021quantum,PhysRevResearch.4.033176,PhysRevApplied.17.064044,khan2021physical,wudarski2023hybrid,PhysRevA.107.042402}. QRC can harness the inherent dynamics of disordered and dissipative quantum systems to learn, predict, and classify various linear and nonlinear temporal tasks, both quantum and classical, including those inspired by human brain functions~\cite{ markovic2020physics, yamamoto2017coherent, tacchino2019artificial}. Essentially, QRC generalizes the classical reservoir learning frameworks such as chaotic liquid state machines and echo-state networks, which are known to significantly reduce the optimization complexity of conventional recurrent neural networks~\cite{lukovsevivcius2009reservoir, cucchi2022hands,tanaka2019recent,gauthier2021next}. In QRC, a stream of inputs interacts with a quantum system (the ``reservoir'') and the system undergoes quantum evolution described by a completely positive and trace-preserving quantum map~\cite{caruso2014quantum}. After an optimal time set by physical parameters, measurements of the reservoir are post-processed with classical learning techniques to form the computing device. This approach can be contrasted with the extensively studied variational quantum algorithms, where the optimization capability scales unfavorably with the number of parameters, limiting scalability and accuracy specially on noisy hardwares~\cite{cerezo2021variational, larocca2024review, mcclean2018barren,PRXQuantum.2.040316,cerezo2023does}. On the contrary, QRC systems can \emph{benefit} from noise and dissipation and are much easier to scale up~\cite{domingo2023taking,olivera2023benefits,sannia2024dissipation,kubota2022quantum}.
\begin{figure}
    \centering
\includegraphics[width=0.85\columnwidth]{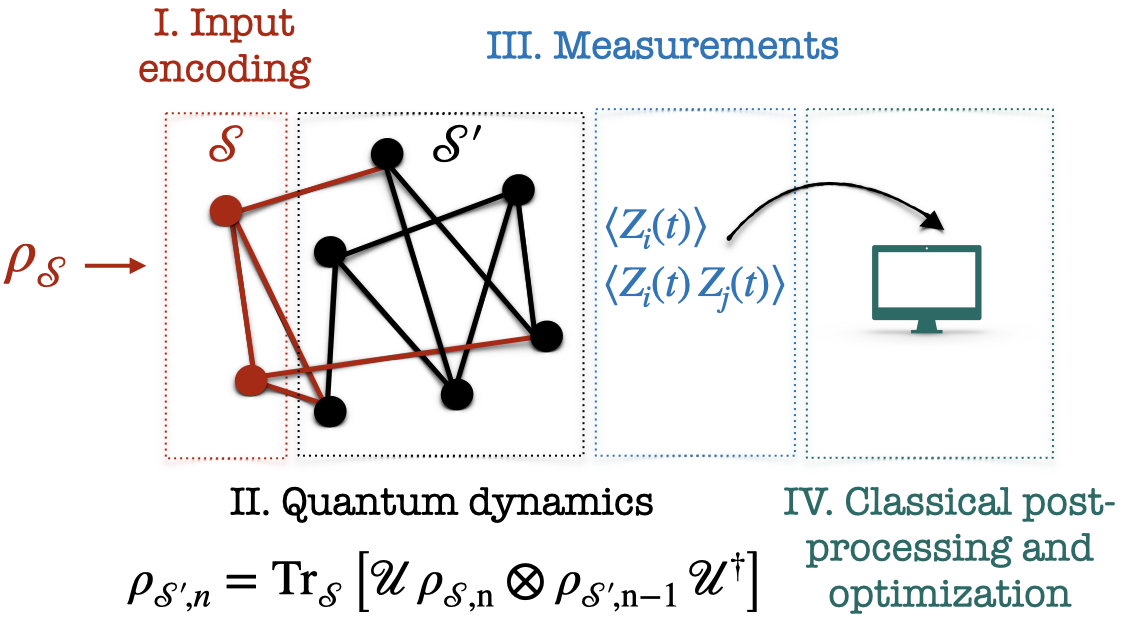}
\caption{An auxiliary system $\mathcal{S}$ is initialized in density matrix $\hat{\rho}_{\mathcal{S},n}$, encoding input data stream at step $n$. $\mathcal{S}$ interacts with the reservoir $\mathcal{S}'$ for the time-scale $J^z\Delta t$ through $\hat{\mathcal{U}}(\Delta t)=e^{-i\hat{\mathcal{H}}\Delta t}$, before the next input is injected. The time-independent Hamiltonian $\hat{\mathcal{H}}$ is defined on a graph with $N$ spins each connected to exactly $k$ random neighbours. Measurement results of the reservoir degrees of freedom $\langle\cdot\rangle$, which are expectation values of spin operators with respect to the current state of the reservoir $\hat{\rho}_{\mathcal{S}',n}$, are then recorded for classical post-processing and optimization purposes. Repeating this for a sequence of temporal data, one can construct a recurrent quantum channel $\hat{\rho}_{\mathcal{S}',n}=\mathcal{L}_{\mathcal{S}}\left(\hat{\rho}_{\mathcal{S'}, n-1}\right)$, capable of learning and emulating various real-world and neurological tasks.}
    \label{fig:reservoir}
\end{figure}

Utilizing disordered, interacting quantum many-body systems as computing reservoirs requires identifying the system setups and parameters which optimize the learning process. To this end, here we study the dependence of learnability performance on the underlying model, in particular, on the connectivity of the spin model and the strength of the interactions and disorder. While previous theoretical studies have focused on either one-dimensional or fully connected models for the quantum reservoir, we study spin models defined on random regular graphs (RRGs), and find that the graph degree together with quantum correlations can significantly impact learnability. Notably, in the fully connected limit we observe that some advantages diminish. This is partly because in this limit the computing reservoirs become ``too effective" at spreading the information contained in the inputs non-locally while the measurements that are used to feed the classical post-processing are local. Moreover, graph connectivity can influence the integrability and chaotic properties. Densely connected random spin models may show a suppression of quantum chaos~\cite{grabarits2024quantum}, which we demonstrate is relevant to our findings.
 
{\emph{Model and dynamics}. We consider the following Hamiltonian of interacting quantum spins on a RRG:
\begin{equation}
    \hat{\mathcal{H}}=\sum_{ij,\alpha}J^{\alpha}_{ij}\,\hat{\sigma}^\alpha_i \hat{\sigma}^\alpha_j  +\sum_{i,\alpha}h_i^{\alpha}\,\hat{\sigma}^{\alpha}_i,
    \label{ising1}
\end{equation}
where $\hat{\sigma}^\alpha_i$ is the Pauli spin-$1/2$ operator at the vertex $i$ and $\alpha\in\{x,z\}$ determines the spin direction. The coupling $J^{\alpha}_{ij}=J^{\alpha}A_{ij}$, where $A_{ij}$ are the elements of the adjacency matrix of a graph  with $N$ spins. The degree $k$ of a vertex (site) is defined as the number of edges connected to that vertex, and a RRG is a graph where all vertices have the same degree and the edges are randomly assigned. Such model Hamiltonian can represent different types of systems, including ion traps, as well as nuclear and electronic spins~\cite{RevModPhys.86.153}. We set \( h^{\alpha}_{i} = h^{\alpha} + \delta^{\alpha}_i \), where \( \delta^{\alpha}_i \in [-\Delta^{\alpha}, \Delta^{\alpha}] \). We also fix \( (J^z, h^z, h^x) = (1, 0, 1) \), and \( \Delta^{z} = 0.2 \). The last term ensures the absence of extra Hamiltonian symmetries. All energy and time-scales are given in terms of $J^{z}$.

We separate the total system into the auxiliary system $\mathcal{S}$ used for data input and the reservoir $\mathcal{S}'$ (see Fig.~\ref{fig:reservoir}). Computation is carried out in steps, as follows: Input data is encoded into the density matrix \(\hat{\rho}_{\mathcal{S}}\) for the subsystem \(\mathcal{S}\), and the initial state of the entire system is given by \(\hat{\rho}_{\mathcal{SS'}} \rightarrow \hat{\rho}_{\mathcal{S}} \otimes \hat{\rho}_{\mathcal{S'}}\), where the reservoir state is defined as \(\hat{\rho}_{\mathcal{S'}} = \mathrm{Tr}_{\mathcal{S}} \left[\hat{\rho}_{\mathcal{SS'}}\right]\). The system evolves unitarily under the Hamiltonian~ Eq.~\ref{ising1} over a time interval \(\Delta t\), and this process is repeated iteratively. After \(n\) input steps, the density matrix of the system \(\hat{\rho}_{\mathcal{SS'}}\) becomes:
\begin{align}
    \hat{\rho}_{\mathcal{SS'}} (n \Delta t) = \hat{\mathcal{U}}(\Delta t) \, \hat{\rho}_{\mathcal{S},n} \otimes\hat{\rho}_{\mathcal{S'}}\left[(n-1) \Delta t \right] \,\hat{\mathcal{U}}^{\dagger}(\Delta t),
\end{align}
where \(\hat{\mathcal{U}}(\Delta t) = e^{-i \hat{\mathcal{H}} \Delta t}\). Here, \(\hat{\rho}_{\mathcal{S},n}\) encodes the \(n\)-th input, and \(\hat{\rho}_{\mathcal{S'}}\left[(n-1) \Delta t \right]\) represents the reservoir state after evolving for \(\Delta t\) following the \((n-1)\)-th input step. The quantum map describing this dynamics is strictly contractive, ensuring fading-memory and convergence in an optimal dynamical regime~\cite{martinez2021dynamical}. This resembles the Stinespring representation of a quantum channel, where the evolution of a physical open quantum system can be described as partial trace of a unitary operation on a composite system in a dilated Hilbert space~\cite{caruso2014quantum}. Alternatively, as a result of consecutive input injections and resetting, this is equivalent to an incoherent process on the auxiliary system~\cite{sannia2024dissipation,olivera2023benefits}.

\begin{figure}[b!]
    \centering
    \includegraphics[width=0.48\columnwidth]{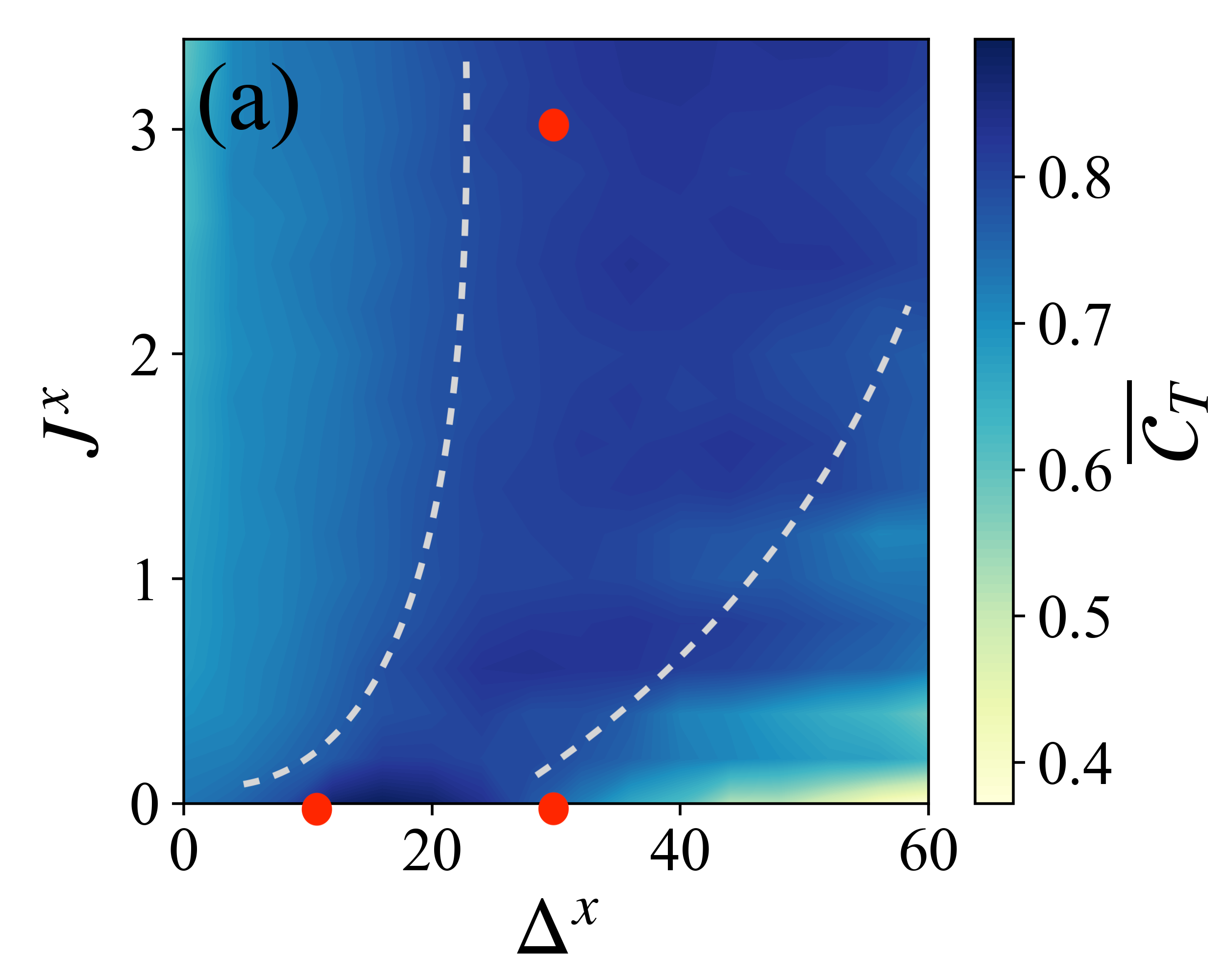}
    \includegraphics[width=0.5\columnwidth]{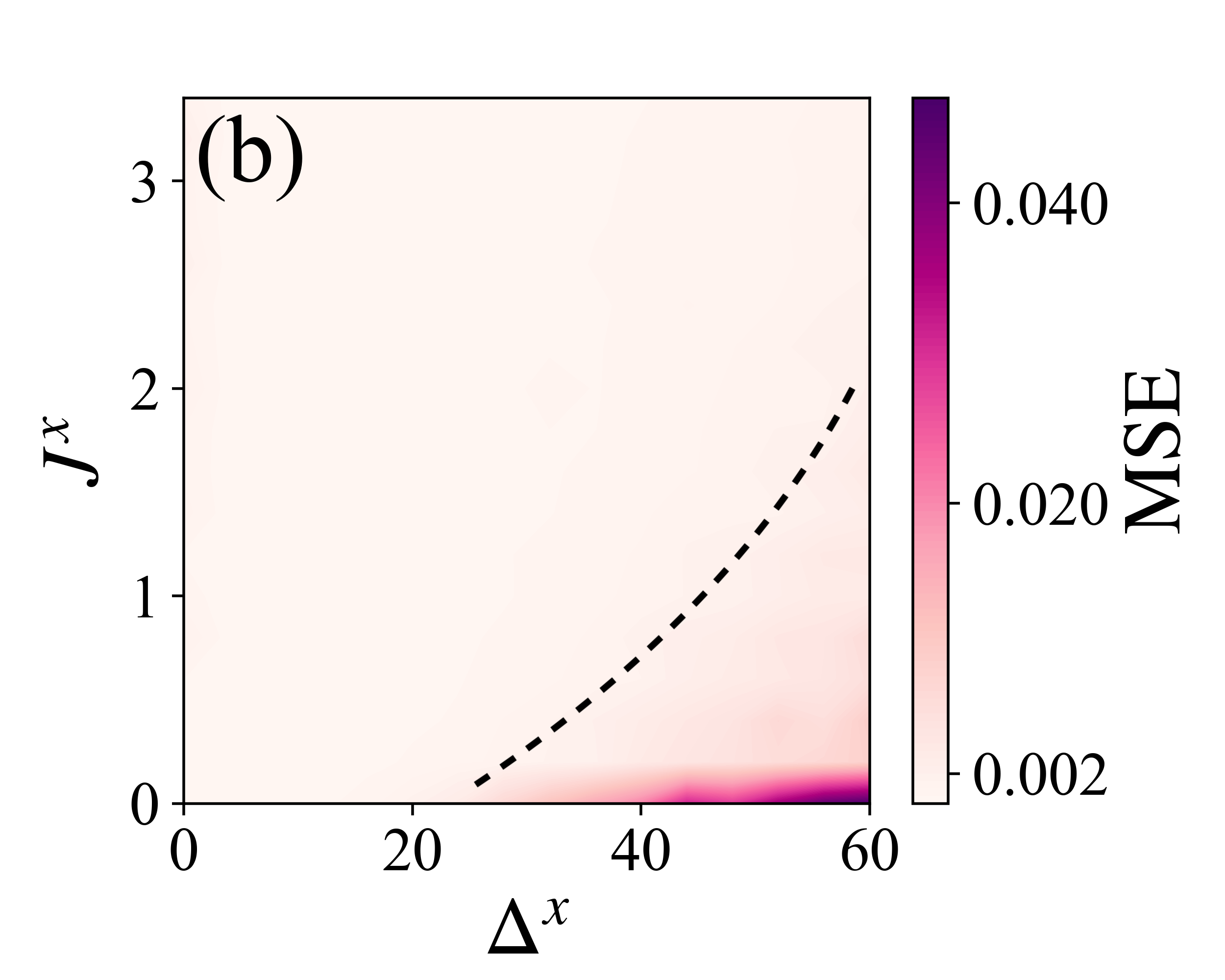}
    \caption{Learning diagrams as a function of disorder $\Delta^x$ and interaction $J^x$.{\textbf{(a)}} Normalized total memory capacity $\overline{\mathcal{C}_{T}}$ for delay time $1\leq\tau\leq6$ and {\textbf{(b)}} MSE for $\tau=1$. Dashed lines approximately mark regions where $\overline{\mathcal{C}_{T}}\geq0.75$ and $\rm MSE\leq2.5\times10^{-3}$. Red dots indicate the places in the diagram where most of the analysis in this work is conducted. Evidently, performance is best at the ``edge of chaos'', when the system is at the cusp of becoming localized. Calculated for $(N,k)=(8,3)$ and $J^{z}\Delta t=3$. } 
    \label{fig:pds}
\end{figure}

\begin{figure*}
    \centering
\includegraphics[width=0.6\columnwidth]{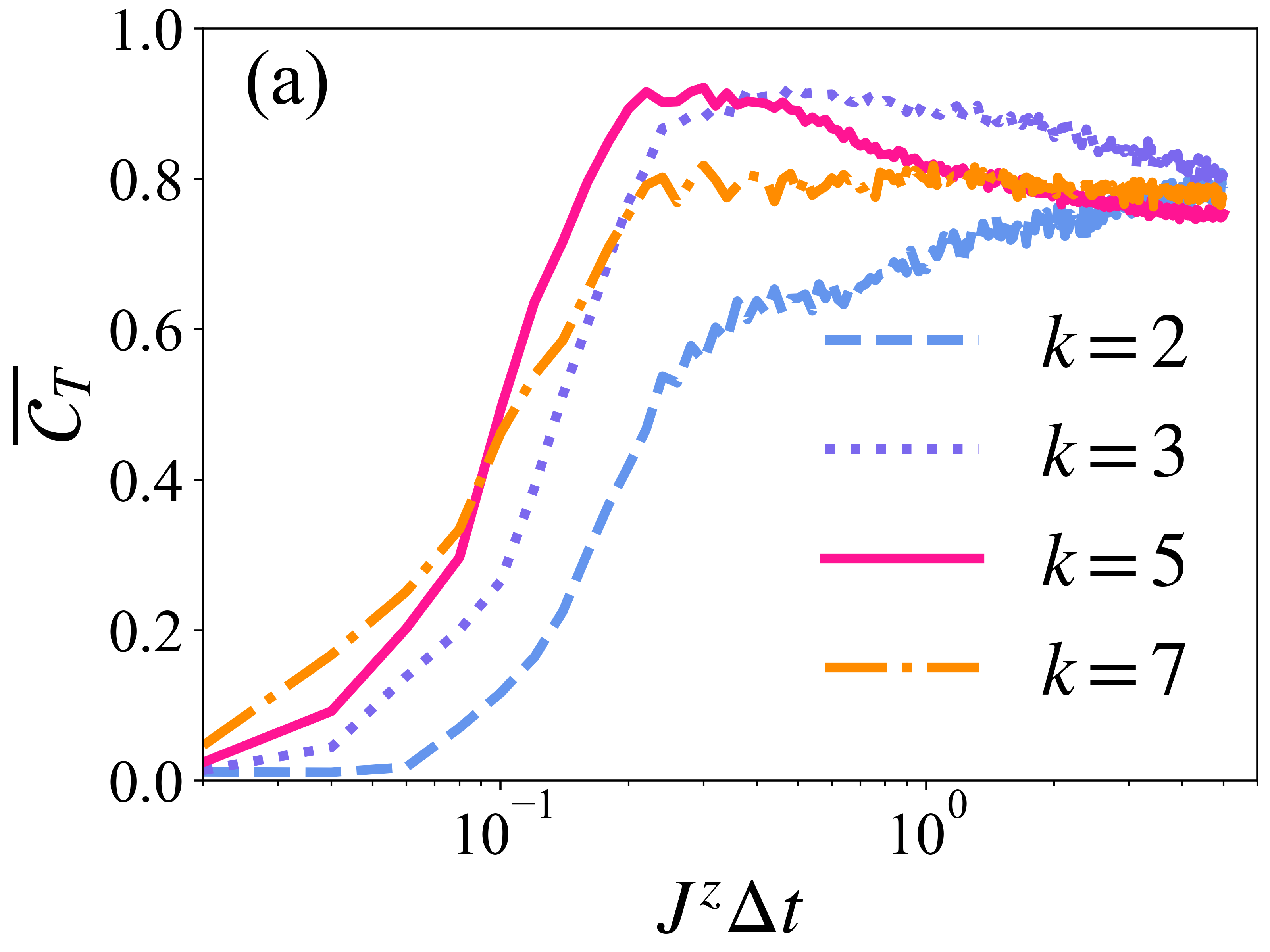}
    \includegraphics[width=0.6\columnwidth]{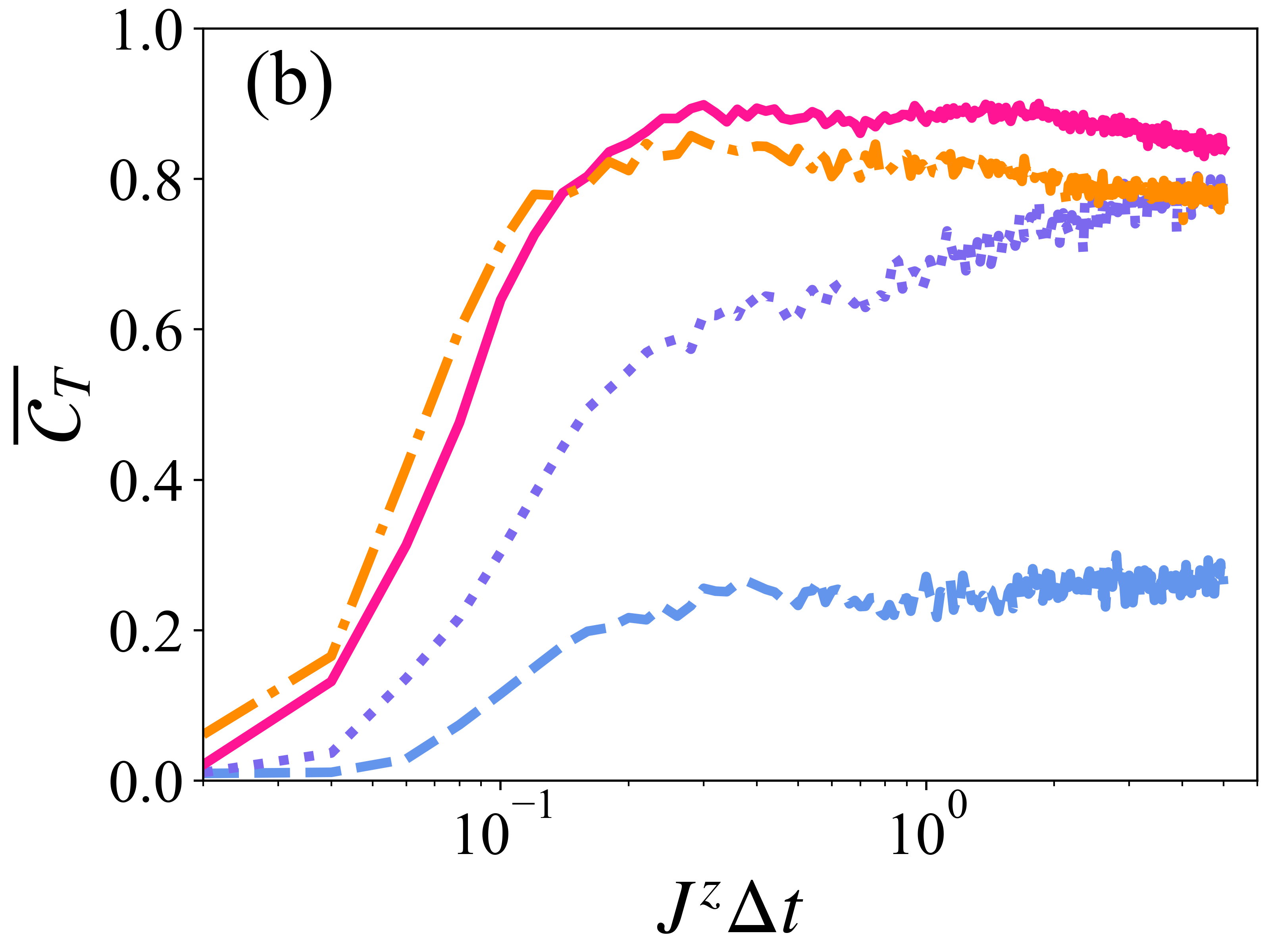}
    \includegraphics[width=0.6\columnwidth]{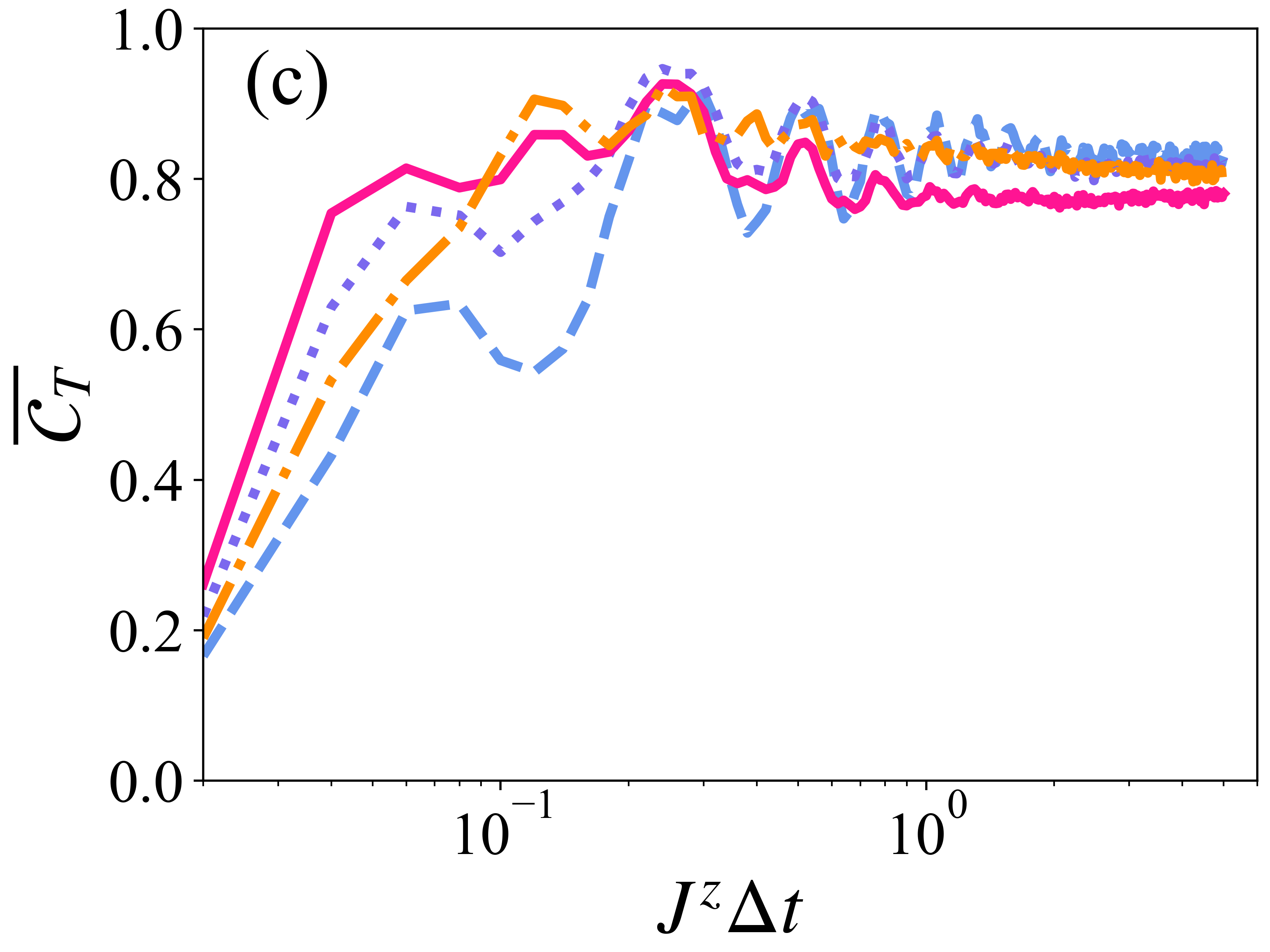}
    \caption{Behavior of the normalized quantum memory capacity $\overline{\mathcal{C}_{T}}$ as a function of the interaction time-scale $J^z\Delta t$. \textbf{(a)} $(\Delta^{x}, J^{x})=(10,0)$, \textbf{(b)} $(\Delta^{x}, J^{x})=(30,0)$. \textbf{(c)} $(\Delta^{x}, J^{x})=(30,3)$. Calculated for $N=8$ and $1\leq\tau\leq6$. Ordering interactions are particularly crucial for rapidly reaching the asymptotic memory limit and optimizing the functionality of low-degree disordered systems.}
    \label{fig:total_mem_vs_dt}
\end{figure*}

{\it Training, Learning and Readout}. The input-output relation of a quantum reservoir supplemented with a classical learning layer can be summarized in a functional form as $\{\overline{y}_{n}\}=\mathcal{F}\left( \{\hat{\rho}_{\mathcal{S},n}\}, \hat{\rho}_{\mathcal{SS'},n}, \mathcal{W}\right)$. Here, $\{\overline{y}_{n}\}$ indicates the set of predictions obtained after classical post-processing. These predictions are derived by minimizing an error measure with respect to a sequence of desired targets ${y_n}$ through a learning process. During training, the measurements of local expectation values $\langle\hat{\sigma}^\alpha_i (n\Delta t)\rangle = {\rm Tr}[\hat{\rho}_{\mathcal{S}'}(n\Delta t)\,\hat{\sigma}^\alpha_i] $, and two-point correlation functions, $\langle \hat{\sigma}^\alpha_i(n\Delta t)\, \hat{\sigma}^\alpha_j(n\Delta t) \rangle$, where $i,j \in \mathcal{S}'$, are used as ``features". These features are combined linearly to fit the desired targets ${y_n}$, and the optimal weights $\mathcal{W}$ are determined by this fitting process. Importantly, we only consider measurements in the computational basis $\alpha=z$ and \textit{do not} apply time-multiplexing techniques~\cite{fujii2017harnessing}, which increases the complexity in number of measurements drastically. Moreover, we do not address the back-action of projective measurements~\cite{mujal2023time,kobayashi2024feedback,yasuda2023quantum}. However, in time-series processing, a full temporal reset of the system is unnecessary. Thanks to the fading memory, only the recent state of the system is relevant for making accurate predictions ~\cite{mujal2023time,chen2019learning}. Therefore, our setup can achieve linear time complexity for such learning tasks, without additional scaling factors from time multiplexing or measuring in multiple bases. To quantify the information retrieval accuracy, we employ the Pearson correlation coefficient $\mathcal{C}_{n}={\rm {cov}}^2(y_n,\bar{y}_n)/\left({\rm{var}} (y_n) ,{\rm{var}} (\bar{y}_{n})\right)$~\cite{tran2021learning,martinez2021dynamical}, where $\bar{y}_n$ denotes the predicted value at step $n$. The coefficient $\mathcal{C}_{n}$ is bounded between 1 and 0, indicating complete or no linear correlation between the predicted and target values, respectively. We also estimate prediction accuracy using the averaged mean-squared error, defined as ${\rm MSE}=\sum_{n}^{N_{L}}(y_n-\bar{y}_n)^2/N_{L}$, where $N_{L}$ is the length of the input data. The reported results are averaged over 50-200 independent realizations of random Hamiltonians, graphs, and input sequences, and the averages are taken over the results of the evaluation stage (See Appendix A for details).

Given this discussion, it is clear that the dynamical properties of the reservoir play a pivotal role in learning. In particular, localization behavior of quantum Hamiltonian models defined on RRGs is directly affected by the connectivity, i.e., the graph degree $k$~\cite{de2020subdiffusion,tikhonov2019critical,kochergin2024robust}. In earlier works it was found that, akin to some classical computation models and for tasks which require both sufficient memory and degree of nonlinearity, QRC systems achieve optimal performance at the \textit{edge of chaos}, or more generally at the vicinity of a (quantum) phase transition~\cite{martinez2021dynamical,tran2021learning, xia2022reservoir,legenstein2007edge,langton1990computation,o2022critical}. Here we will study whether this is also the case in our model.
 
{{\emph{Quantum Tomography and Memory}}.} As our first example, we study a linear quantum task. Consider the following family of bipartite input quantum states, known as Werner states~\cite{werner1989quantum, uola2020quantum}
\begin{align}
\hat{\rho}_{\rm W}(\eta,d,t)= \frac{d-1+\eta(t)}{d-1} \frac{\hat{\mathbb{I}}}{d^2}- \frac{\eta(t)}{1-d} \frac{\hat{\mathcal{V}}}{d} ,
\end{align}
given as a mixture of a swap operator ${\hat{\mathcal{V}}}$ and a maximally mixed state $\hat{\mathbb{I}}$, with the mixing parameter $\eta$. The swap operator is defined as ${\hat{\mathcal{V}}}(\ket{\Psi}\otimes\ket{\Phi})=\ket{\Phi}\otimes\ket{\Psi}$, exchanging the states of a bipartite quantum state. $d$ is dimension of the input and here indicates the number of ancillary qubits. We consider two-qubit input states and can write $\hat{\rho}_{\rm W}(\eta',t)= \frac{1}{4} (1-\eta'(t)) \hat{\mathbb{I}}- \eta'(t) \hat{\rho}_{\rm B}$, with $\hat{\rho}_{\rm B}=\ket{\Phi}\bra{\Phi}$ and $\ket{\Phi}=(\ket{\uparrow\downarrow}-\ket{\downarrow\uparrow})/\sqrt{2}$ a singlet Bell state~\cite{li2021information,PhysRevLett.92.177901}. With $0\leq\eta'\leq1$, $\hat{\rho}_{\rm W}$ is entangled for $\eta'>1/(d+1)$ and separable otherwise~\cite{PhysRevA.59.4206, RevModPhys.81.865}. For a given unitary $\hat{\mathcal{U}}$, this family of bipartite quantum states (by definition) satisfy $\hat{\rho}_{\rm W}=\hat{\mathcal{U}}\otimes\hat{\mathcal{U}}\,\hat{\rho}_{\rm W}\,\hat{\mathcal{U}}^{\dagger}\otimes \hat{\mathcal{U}}^{\dagger}$; a property which is of practical interest in quantum steering and communication protocols~\cite{PhysRevLett.98.140402}. A large family of quantum states, including both Werner and isotropic states, can be expressed in a similar manner, where only a single mixing parameter can characterize the state uniquely. A high fidelity temporal learning of the mixing parameter $\eta(t)$ is thus a proxy to learning dynamical evolution of quantum correlations of input states. Generalization of the described learning scheme to higher-dimensional inputs is straightforward and only requires the ability to encode such states. To evaluate the linear memory capacity, we set the learning target to recovering previous inputs, $y_{n,\tau}\equiv y\left(n\Delta t-\tau\Delta t\right)=\eta' \left(n\Delta t-\tau\Delta t\right)$. The total memory capacity for delayed construction of previous inputs is defined as ${\mathcal{C}}_T=\sum_{n,\tau} \mathcal{C}_{n,\tau}$, with $\tau\geq0$ an integer specifying the \textit{delay time} and $\mathcal{C}_{n,\tau}$ the Pearson correlation coefficient for $y_{n,\tau}$. We note that $\mathcal{C}_{n,\tau}\to 0$ when $\tau\to \infty$, and for a finite delay time we can normalize the averaged total memory $\overline{\mathcal{C}}_T=\mathcal{C}_T/\tau_{\rm max}$.

{\it Optimal Learning Regimes.} Figure~\ref{fig:pds} displays a learning diagram for $\overline {\mathcal{C}}_T$ and MSE of the predicted values in the $\Delta^x-J^{x}$ plane. The optimal learning regime for our model occurs at around the boundary of chaotic-localized phase transitions. Notably, for a given interaction timescale $J^z\Delta t$ and for a graph with a fixed degree $k$, the addition of interactions $\hat{\sigma}^{x}_i\,\hat{\sigma}^{x}_j$ in the disordered regimes is advantageous to both memory and also short-term predication accuracy. As we will numerically establish later, this term represents an entangling and delocalizing interaction. The behavior of memory capacity as a function of the $\mathcal{S}-\mathcal{S}'$ interaction timescale $J^z\Delta t$ is shown for selected points in Fig.~\ref{fig:total_mem_vs_dt}. In certain dynamical regimes, there can be a window where the largest memory performance is achieved, in accordance with previous studies~\cite{fujii2017harnessing,tran2021learning}. As  disorder increases, only high-degree reservoirs exhibit rapid initial growth of memory, followed by saturation in the long-time limit. As depicted in Fig.~\ref{fig:total_mem_vs_dt}(b), for $k=2,3$ the memory capacity exhibits a slow behavior, hinting at the slow propagation of information in the strongly disordered regime. Adding the ordering interactions,  as illustrated in Fig.~\ref{fig:total_mem_vs_dt}(c), recovers the fast growth of the quantum memory capacity and allows the system to reach the optimal possible performance for all degrees; we attribute this to their delocalizing effect.

The main features discussed above are also evident in Fig.~\ref{fig:fig4}, where we additionally observe that, for the specified time intervals, higher connectivity elevates the memory capacity and improves the short-term prediction errors. However, the most optimal learning regime, in terms of both memory and accuracy, is achieved by tuning moderate interactions and computing on graphs with an intermediate $k/N$ ratio. While reservoirs defined on higher degree graphs can evade localization in the presence of stronger disorder, it becomes exceedingly difficult to extract the non-locally hidden inputs information through only (quasi-) local measurements when $k\to N-1$. In such cases, retrieving the inputs information in a chosen measurement basis possibly requires measuring higher-order correlation functions of the form $\langle{\hat{\sigma}}^{\alpha}_i{\hat{\sigma}}^{\alpha}_j\cdots{\hat{\sigma}}^{\alpha}_N\rangle$, which should be useful as extra learnable features in the training stage. Importantly, the drop in performance can also be attributed to changes in quantum chaotic behavior. In Appendix B we demonstrate that in the densely connected limit, similar to recent findings~\cite{grabarits2024quantum}, the level spacing ratio shows signs of integrability. 

\begin{figure}
    \centering
    \includegraphics[width=0.8\columnwidth]{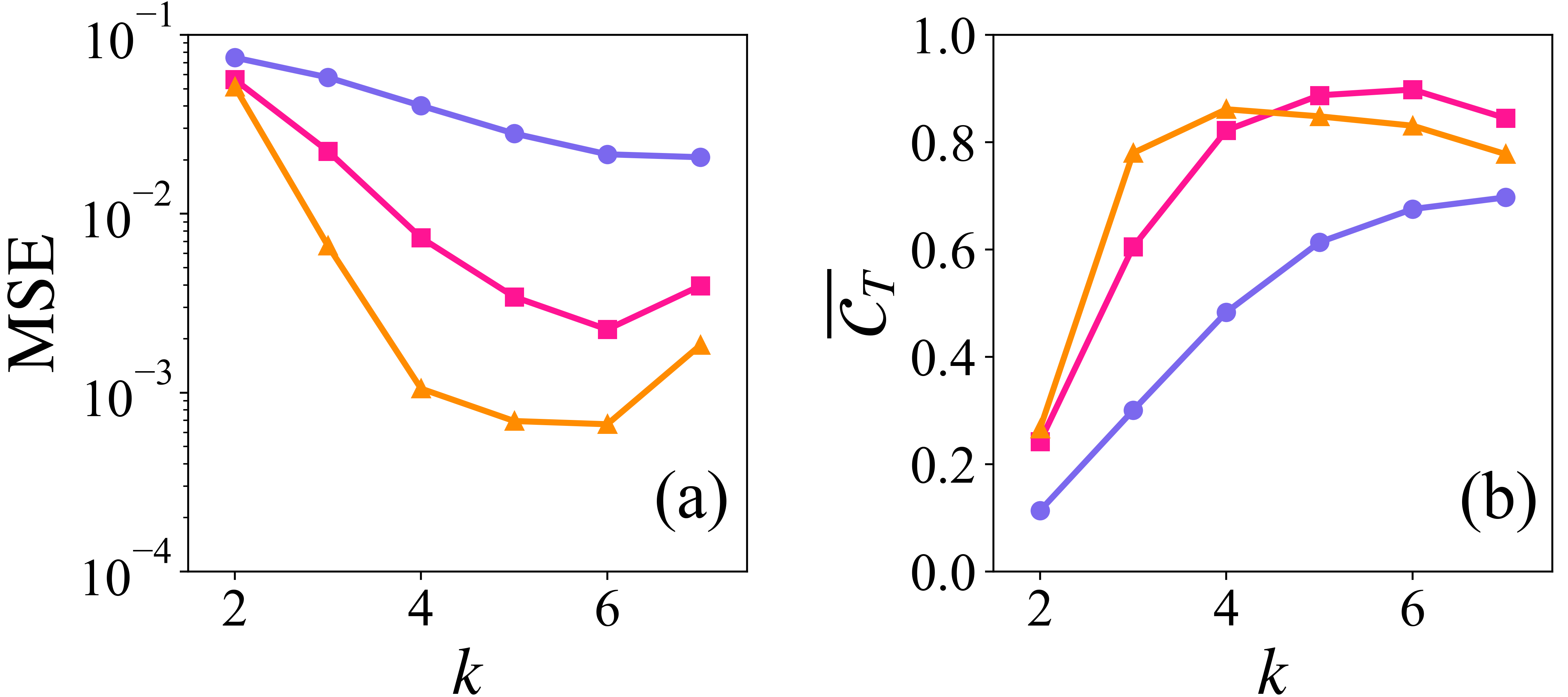}
    \includegraphics[width=0.8\columnwidth]{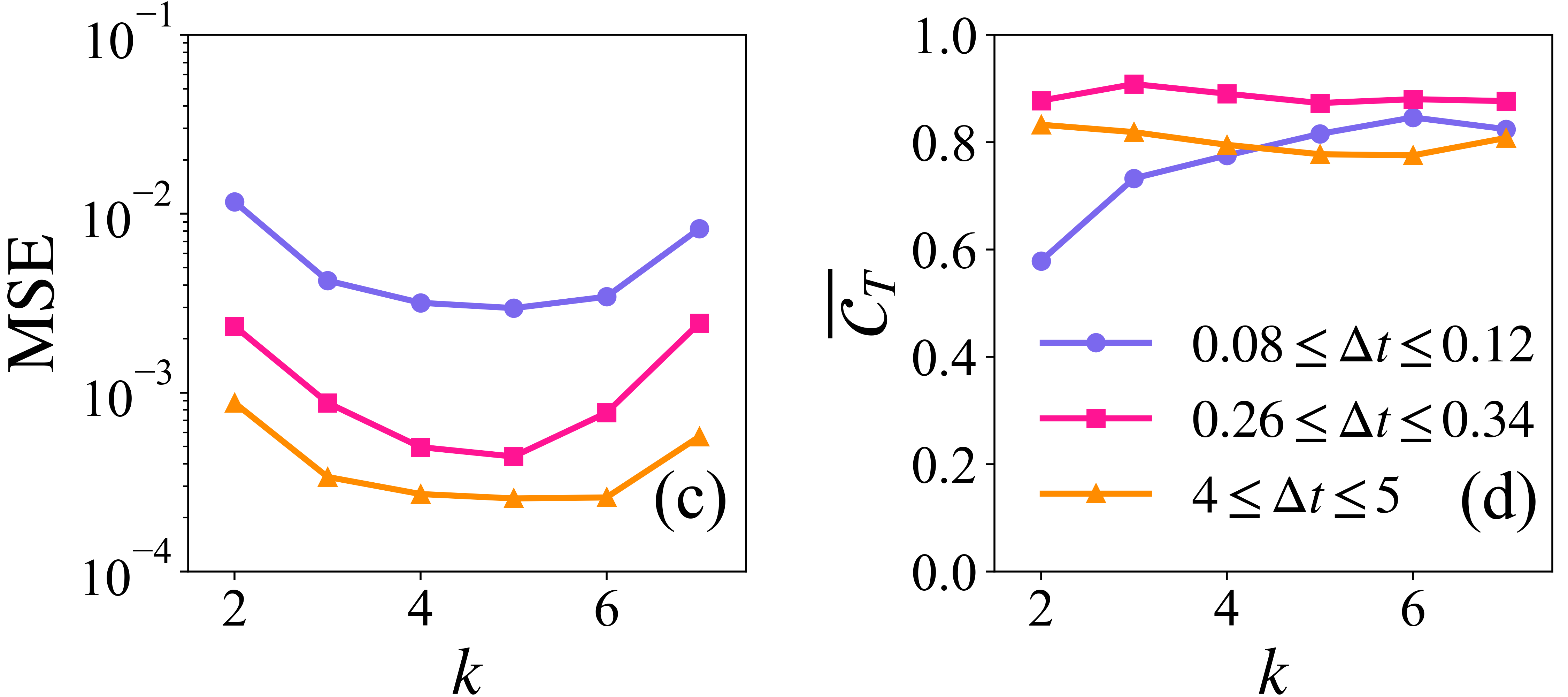}
    \caption{Time-averaged MSE and normalized total memory capacity $\overline{\mathcal{C}}_T$ as a function of the graph degree $k$ for $N=8$. Plotted for \textbf{(a)-(b)} $(\Delta^{x}, J^{x})=(30,0)$ and 
    \textbf{(c)-(d)} $(\Delta^{x}, J^{x})=(30,3)$. Memory and MSE are reported for $1\leq\tau\leq6$ and $\tau=1$, respectively. Apart from highlighting the role of $\Delta t$ in our setup, disordered reservoirs achieve the most optimal computing performance, in terms of both accuracy and memory, on intermediate-degree graphs and in presence of quantum interactions.}
    \label{fig:fig4}
\end{figure}

{\it Correlation and Entanglement.} The performance of quantum reservoirs can be related to fundamental and physical measures, such as degree of ``quantumness", information scrambling and dynamical correlations~\cite{gotting2023exploring,senanian2023microwave,PhysRevResearch.4.033007, kutvonen2020optimizing, xia2022reservoir, domingo2023quantum,kobayashi2023quantum}. Here we study an alternative version of the correlation operator introduced in Refs.~\cite{PhysRevX.10.041024,PhysRevResearch.6.013243}, and define the auxiliary system-reservoir correlation as $\hat{\chi}(t)=\hat{\rho}_{\mathcal{SS}'}(0)-\hat{\rho}_{\mathcal{SS}'}(t)$, where $\hat{\rho}_{\mathcal{SS}'}(0)=\hat{\rho}_{\mathcal{S}}(0)\otimes\hat{\rho}_{\mathcal{S}'}(0)$ and $\hat{\rho}_{\mathcal{SS}'}(t)=\hat{\mathcal{U}}(t)\,\hat{\rho}_{\mathcal{SS}'}(0)\,\hat{\mathcal{U}}^{\dagger}(t)$. We numerically calculate $\lVert\hat{\chi}(t)\rVert$, with $\lVert\cdot\rVert$ the Hilbert-Schmidt norm. This probe measures the degree of total correlation introduced by the dynamics between the initially unentangled $\mathcal{S}$ and $\mathcal{S}'$, and here it can also be interpreted simply as a distance measure. We set $\hat{\rho}_{\mathcal{S}}=\hat{\rho}_{W}$ with some random $\eta'$ and average over different realization of initial states and disordered Hamiltonians. As shown in Fig.~\ref{fig:corr_logneg}(a), initially $\lVert\hat{\chi}(0)\rVert=0$, indicating no correlation between auxiliary and reservoir degrees of freedom. As time passes by, $\lVert\hat{\chi}(t)\rVert$ displays an initial growth within the time-scale $\tau\propto 1/\Delta^x$ for $\Delta^x/J^z>1$. Notably, with an optimal disorder level, composite system can avoid localization while leveraging chaotic dynamics for rapid and enhanced developments of correlations. While being sensitive to certain forms of information localization and the spectral properties of the underlying Hamiltonian, this measure lacks the ability to distinguish between quantum and classical correlations.
\begin{figure}[t]
\centering
\includegraphics[width=0.49\columnwidth]{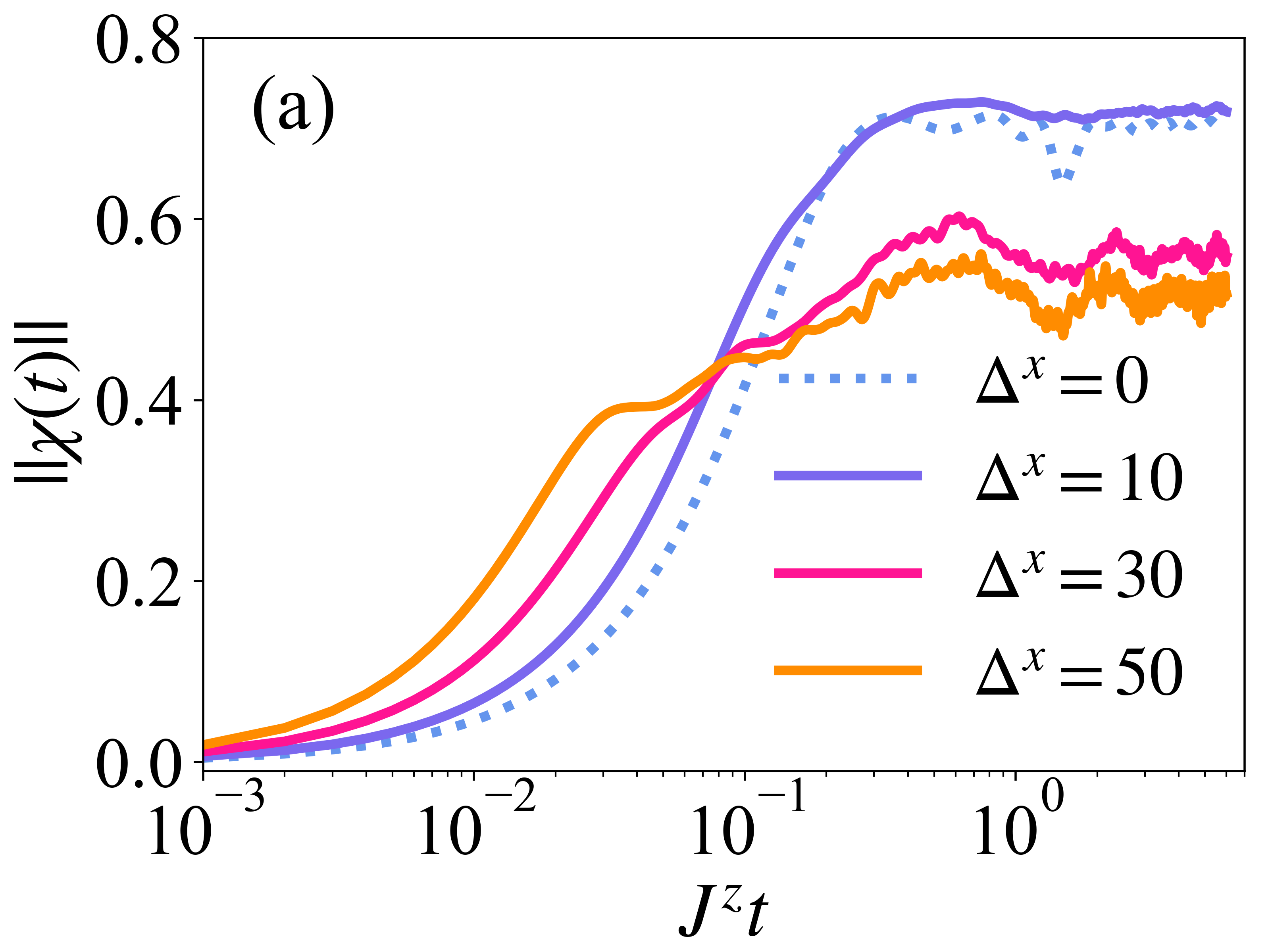}\includegraphics[width=0.49\columnwidth]{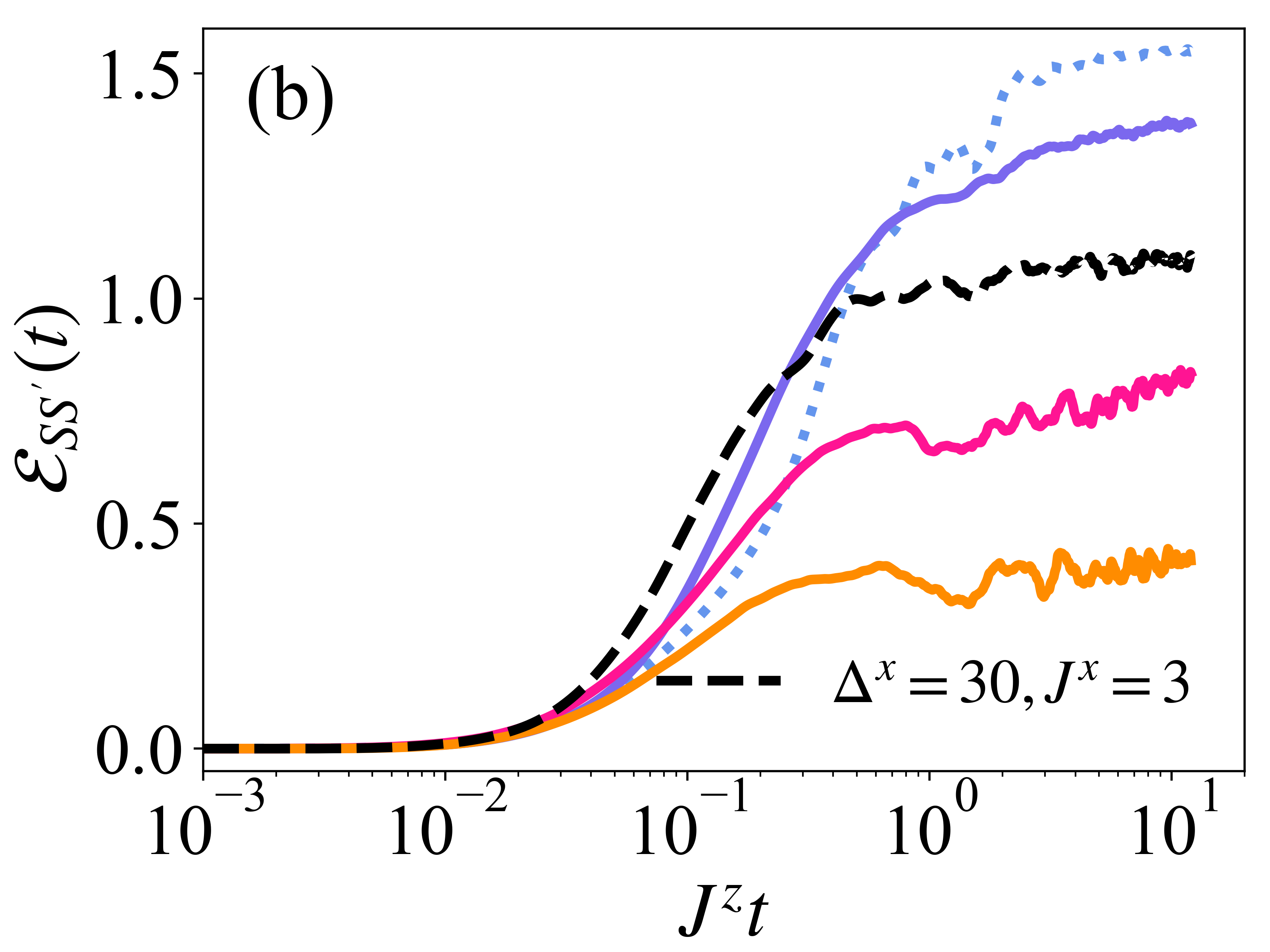}
\caption{\textbf{(a)} Norm of the auxiliary system-reservoir correlation $\lVert\hat{\chi}(t)\rVert$ at various disorder strengths $\Delta^x$, plotted for $J^x=0$. Controlled disorder can aid in creating complex and richer dynamics. \textbf{(b)} Dynamical logarithmic negativity $\mathcal{E}_{{SS}'}(t)$ between auxiliary system $\mathcal{S}$  and the reservoir $\mathcal{S}'$. Here $(N,k)=(8,3)$. Entangling quantum correlations are crucial for avoiding localization and improving learning performance.}
\label{fig:corr_logneg}
\end{figure}

A more useful measure of the dynamical quantum correlation build-up is the mixed-state entanglement between the auxiliary system $\mathcal{S}$ and the reservoir $\mathcal{S}'$. We characterize this here by the logarithmic negativity defined as $\mathcal{E}_{{SS}'}=\log_2\lVert\hat{\rho}^{T_{\mathcal{S}}}_{\mathcal{SS}'}\rVert$, where $T_{\mathcal{S}}$ indicates partial transpose with respect to $\mathcal{S}$ and $\lVert\cdot\rVert$ is the trace norm~\cite{RevModPhys.80.517, PhysRevLett.77.1413, RevModPhys.81.865,PhysRevA.65.032314, PhysRevLett.95.090503}. Finite logarithmic negativity quantifies entanglement cost and entanglement of distillation~\cite{audenaert2003entanglement}. The dynamical behavior of $\mathcal{E}_{{SS}'}$ for different disorder strengths is shown in Fig.~\ref{fig:corr_logneg}(b). The onset of localization is reflected in the slow logarithmic growth of entanglement, in contrast to the chaotic regime with a volume-law entanglement scaling~\cite{abanin2019colloquium}. As can be seen, adding the ordering interactions $\hat{\sigma}^x_i\hat{\sigma}^x_j$ in the large disorder regime recovers the fast growth and produces strong entanglement at short times. The presence of quantum correlations can in turn improve the memory capacity and learnability accuracy of disordered reservoirs, as observed in the previous section. 

{{\emph{Classical Logical Multitasking}}.} To showcase the ability of our spin reservoir in performing nonlinear tasks, we now consider classical logical multitasking~\cite{bravo2022quantum}. Given two independent sequences of binary inputs, the network tries to simultaneously learn how to $\mathtt{AND}$, $\mathtt{OR}$ and $\mathtt{XOR}$ them. We set the state of each input spin to $\hat{\rho}_{n} = (1-\eta_n)\ket{\uparrow}\bra{\uparrow} + \eta_n\ket{\downarrow}
\bra{\downarrow}$ with $\eta_n\,\in\{0,1\}$ encoding the input bits. Figure~\ref{fig:corr_multitask} displays the accuracy of the learned operations as the disorder and connectivity are varied. The $\mathtt{XOR}$ operation is not linearly separable in the two-dimensional input space, and shows higher sensitivity to disorder. However, adding moderate interactions recovers the maximal performance in most regimes, consistent with earlier observations (see Fig.~\ref{fig:corr_multitask}(a)). This supports the expectation that excessive local disorder, in the absence of quantum interactions, can quickly undermine nonlinear information processing capabilities~\cite{martinez2021dynamical}. Remarkably, as shown in Fig.~\ref{fig:corr_multitask}(b), the critical disorder strength $\Delta^x_c$ for the accuracy of the $\mathtt{XOR}$ to fall just below $\approx 0.7$ displays an almost linear dependence on the graph degree, offering a practical tool to control the performance of QRC systems.
\begin{figure}[t]
    \centering
    \includegraphics[width=0.49\columnwidth]{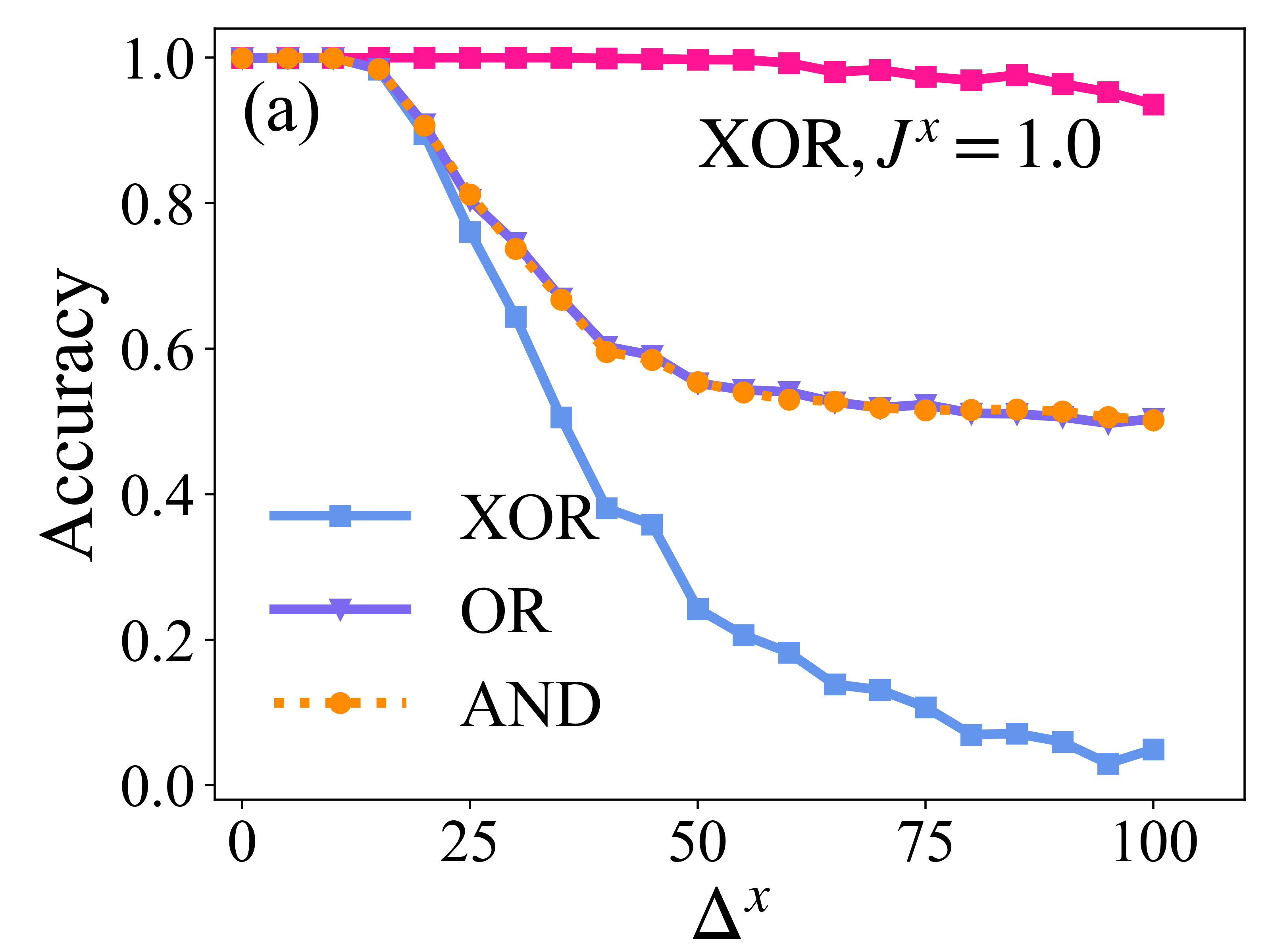}
    \includegraphics[width=0.49\columnwidth]{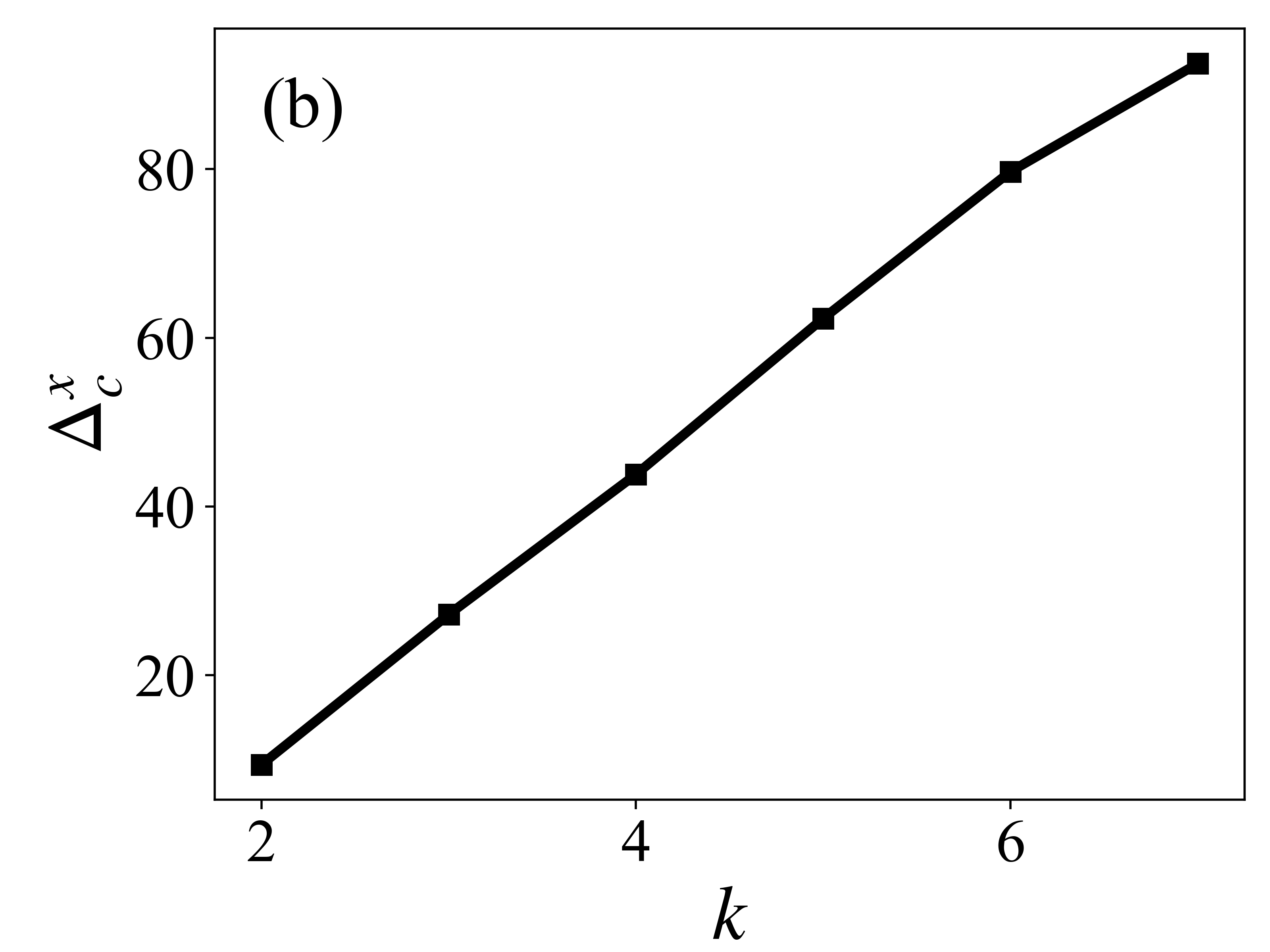}
    \caption{\textbf{(a)} Accuracy of the logical multitasking as a function of disorder strength $\Delta^x$ with $J^x=0$, shown for $(N,k)=(8,3)$ and $J^z\Delta t=3$. Strong disorder degrades the accuracy. \textbf{(b)} Critical disorder strength $\Delta^x_c$ for the $\mathtt{XOR}$ operation as a function of graph degree $k$, with the threshold accuracy set to $\approx 0.7$, calculated for $J^x=0$. Higher-degree graphs are more resistant to disorder. }
    \label{fig:corr_multitask}
\end{figure}

{{\emph{Summary and Discussion.}}} We have introduced a many-body spin reservoir defined on RRGs and evaluated its learning capabilities for various tasks. Our findings demonstrate that, in an optimal dynamical regime, given in terms of disorder, interactions, and connectivity, our model captures the key properties of a high-performing quantum reservoir without requiring time-multiplexing. Our work paves the way towards designing practical analog QRC platforms by linking their performance to fundamental physical and geometrical properties. Our results motivate further research on the impact of geometry and quantum chaotic properties on the performance of quantum learning algorithms. In experimental realizations, utilizing spatial-multiplexing for small quantum systems is a practical option~\cite{nakajima2019boosting}. Possibly, a few random measurements can provide enough information to perform a given learning task with high fidelity~\cite{elben2023randomized, elben2020mixed,PhysRevResearch.6.013211}. Additionally, it would be informative to study how different (symmetry) classes of unitary operations~\cite{domingo2023optimal} and non-trivial electronic topology~\cite{kobayashi2024quantum,bahri2015localization} affect the learning performance. Finally, while here we focused on supervised learning, it is worth exploring the potential for unsupervised approaches for classification tasks.

{{\emph{Acknowledgements.}}} We wish to thank Alexander Balanov, Juan Sebastian Totero Gongora, Sergey Savielev, and Alexandre Zagoskin for useful discussions. We acknowledge the support of the UK Research and Innovation (UKRI) Horizon Europe guarantee scheme for the QRC-4-ESP project under grant number 101129663 and the Academy of Finland through its QTF Center of Excellence program (Project No. 312298).

\bibliography{refs.bib}
\bibliographystyle{apsrev4-1}

\appendix
\section{Appendix A: Details of calculations} 
Here, we provide some details of our numerical calculations. As usual~\cite{fujii2017harnessing}, a number of initial steps are discarded to eliminate transient effects. Specifically, we discard $N_{\rm transient}=600-800$ steps. We use $N_{\rm train}=1000-2000$ steps for training, where each ``training steps'' refers to a point in time where measurements are used to update the model's parameters by comparing the predicted outputs to the target values. Finally, $N_{\rm test}=100-200$ steps are reserved for testing the performance of the trained model. The training data is stored in a matrix of size $N_O\times N_{\rm train}$, where the entry $(i,j)$ represents the $i$-th observable measured at the $j$-th training step. Here, $N_O=N_{\mathcal{S}'}\times(N_{\mathcal{S}'}+1)/2$, with $N_O$ being the total number of measured observables and $N_{\mathcal{S}'}$ the number of spins in the reservoir. We only consider connected graphs, i.e., graphs without isolated subgraphs. 

To emulate encoding errors and avoid overfitting in the training process regarding the memory task studied in the main manuscript, we add small initial noise to inputs by setting ${\eta'}(t)\to\eta'(t)\pm\delta^{\eta}$ with $\delta^{\eta}\in [0, 0.02]$, and rescale properly. For the supervised learning, we utilize the ridge linear regression with appropriate regularization strength (of the order of $10^{-3}-10^{-4}$). For the classification of the logical multitasking, we use support vector machines~\cite{RevModPhys.91.045002}. In this case we nominate a nonlinear Radial Basis Function kernel of the form $\mathcal{K}(X,X')=\exp({-\mathcal{D}(X,X')^2/2l^2)}$. Here $\mathcal{D}(X,X')$ is the Euclidean distance and we set $l=1$. Note, however, that the value of $l$ can vary during hyperparameter tuning. The accuracy for this task is calculated by comparing the binary predictions with the actual inputs. Since the lower bound for predicting a random binary sequence is, on average, $1/2$, we subtract this value to set the reference accuracy to zero, and then multiply by $2$, ensuring it falls within the interval $[0,1]$.

\section{Appendix B: Additional results}

In this section we present supplementary data for the previously unexplored regions of the learning diagrams related to the memory task. Fig.~\ref{fig:fig_app_1}(a) displays the memory capacity in the absence of disorder and interactions $(\Delta^x, J^x)=(0,0)$. In this limit the model reduces to a transverse field Ising model with broken $\mathbb{Z}_2$ symmetry $\prod_i \hat{\sigma}^x_i$:

\begin{equation}
    \hat{\mathcal{H}}=\sum_{ij}J^{z}_{ij}\,\hat{\sigma}^z_i \hat{\sigma}^z_j +\sum_{i}h_i^{x}\,\hat{\sigma}^{x}_i+\sum_{i}h_i^{z}\,\hat{\sigma}^{z}_i.
    \label{ising1}
\end{equation}
We set $ J^{z} = h^{x} = 1 $. Additionally, we define $ h_i^{z} = h^z + \delta_i^z $, where $ h^z = 0 $ and $ \delta_i^z $ represents a set of independent random values chosen from the interval $ \delta_i^z \in [-0.2, 0.2] $.
 Note that since the model is defined on a random graph, interaction terms $ \propto \hat{\sigma}^z_i \hat{\sigma}^z_j $ are also in general random.

Interestingly, in this regime, memory does not appear to improve with increasing graph degree. As mentioned in the main text, this trend is partly due to non-local nature of the encoding. Consistent with the results presented in the main text, memory performs poorly in the densely connected limit and additionally exhibits an anomalous slow dynamics. Crucially, when compared to Fig.~\ref{fig:total_mem_vs_dt}(a) of the main text with $(\Delta^x, J^x)=(10,0)$, it becomes evident that in this limit, random field disorder is essential in regulating chaotic properties and achieving larger memory capacity more quickly. Furthermore, we have calculated the level spacing ratio $r_n={\rm min}[\delta_n,\delta_{n+1}]/{\rm max}[\delta_n,\delta_{n+1}]$, where $n$ labels the sorted eigenvalues and $\delta_n=E_{n+1}-E_n$. The mean level spacing $\langle r\rangle$, averaged over energies, different disordered Hamiltonians and graph realizations, tends to the limiting values $\approx 0.39$ and $\approx 0.53$ for localized (integrable) and ergodic (chaotic) phases, respectively~\cite{martinez2021dynamical}. Notably, as shown in Fig.~\ref{fig:fig_app_1}(b) for $(\Delta^x, J^x)=(0,0)$, the averaged level spacing ratio $\langle r\rangle$ is reduced in the limit of all-to-all connectivity, showing signs of integrability similar to the results of Ref.~\cite{grabarits2024quantum}. At finite disorder, the ratio can as well initially increase as a function of the graph degree $k$, highlighting the delocalizing effects of higher connectivity. In other words, low-degree graphs are more prone to localization in the presence of randomness. Interactions can also induce a tendency to chaotic behavior in strongly disordered regimes. As can be seen, densely connected graphs tend to exhibit a breakdown of chaos in almost all regimes. A careful investigation of these behaviors requires finite-size scaling, which we leave for future studies. 

\begin{figure}[h]
    \centering
\includegraphics[width=0.49\columnwidth]{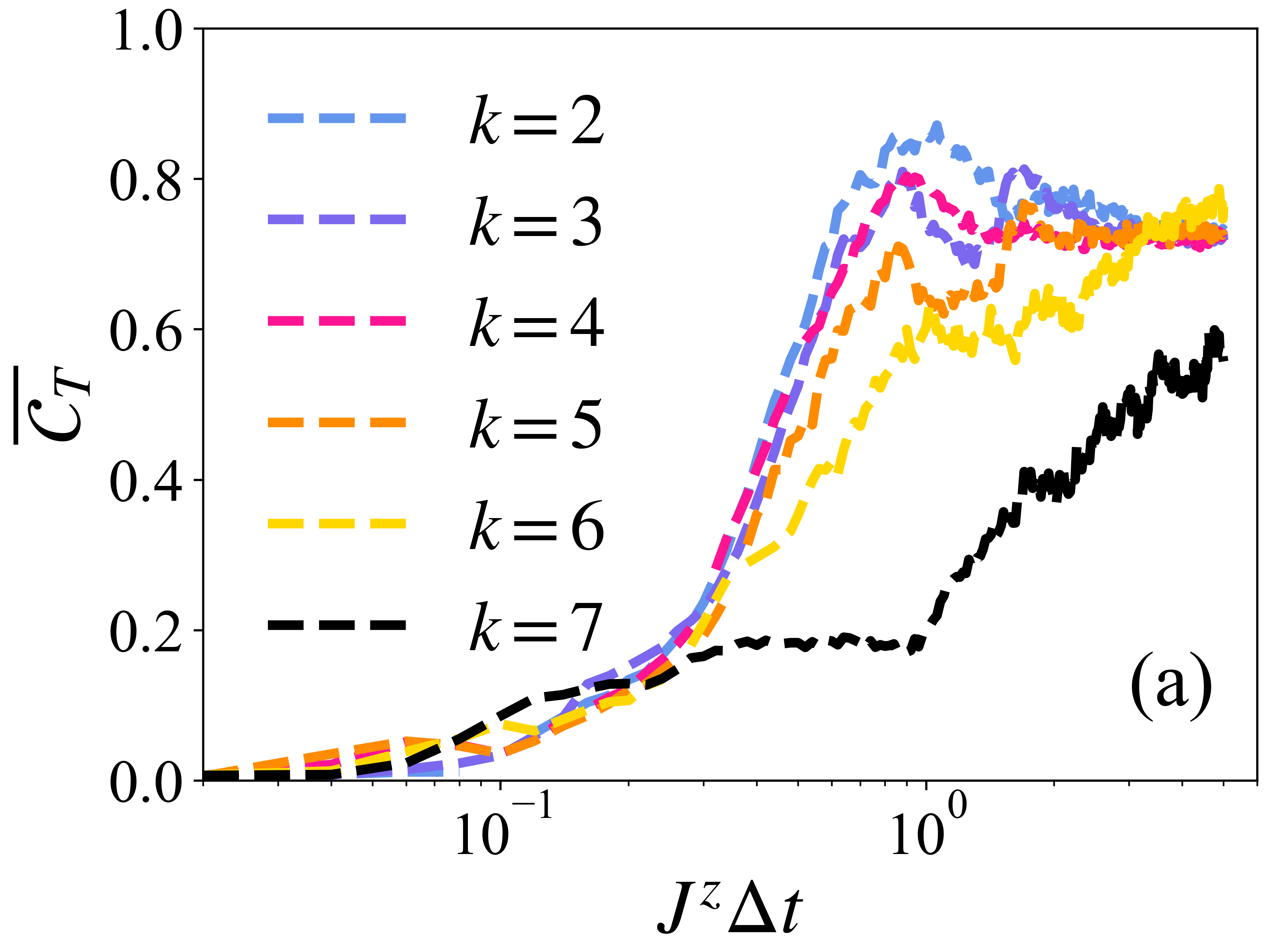}
\includegraphics[width=0.49\columnwidth]{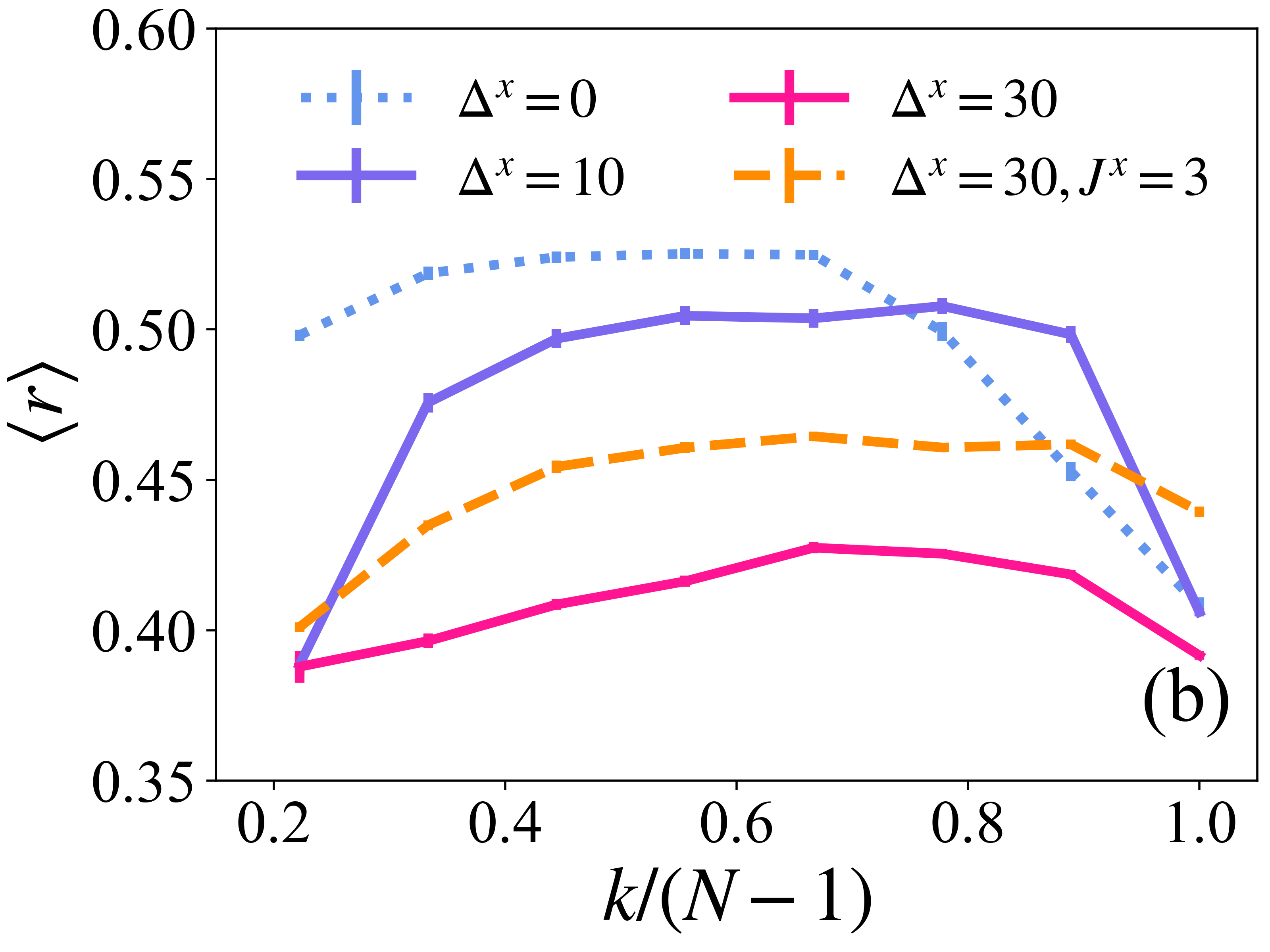}
    \caption{\textbf{(a)} Normalized total memory capacity $\overline{\mathcal{C}}_T$ for $N=8$ and $(\Delta^x, J^x)=(0,0)$. \textbf{(b)} Mean level spacing ratio $\langle r\rangle$ for $N=10$, averaged over 100 independent realizations. }
    \label{fig:fig_app_1}
\end{figure}

To reveal the dynamical effects of connectivity on propagation of quantum correlations, we calculate the logarithmic negativity $\mathcal{E}_{{SS}'}$ for the extreme cases of $k=2$ and $k=N-1$ with $N=8$. In Fig.~\ref{fig:fig_app_2}(a) we plot $\mathcal{E}_{{SS}'}$ for $k=2$. Compared to the the case $k=3$ studied in the main text, disorder can strongly suppress the development of the quantum correlations between the input subsystem and the reservoir. For $k=N-1$ and in the absence of disorder and interactions, the negativity exhibits an anomalous dynamical behavior, as shown in Fig.~\ref{fig:fig_app_2}(b). This reveals the dynamical effects of the emergent integrability, resulting in a very slow propagation of quantum correlations in this limit. It can be easily verified (numerically) that these behaviors are independent of the choice of initial states.

\begin{figure}[h]
    \centering
\includegraphics[width=0.49\columnwidth]{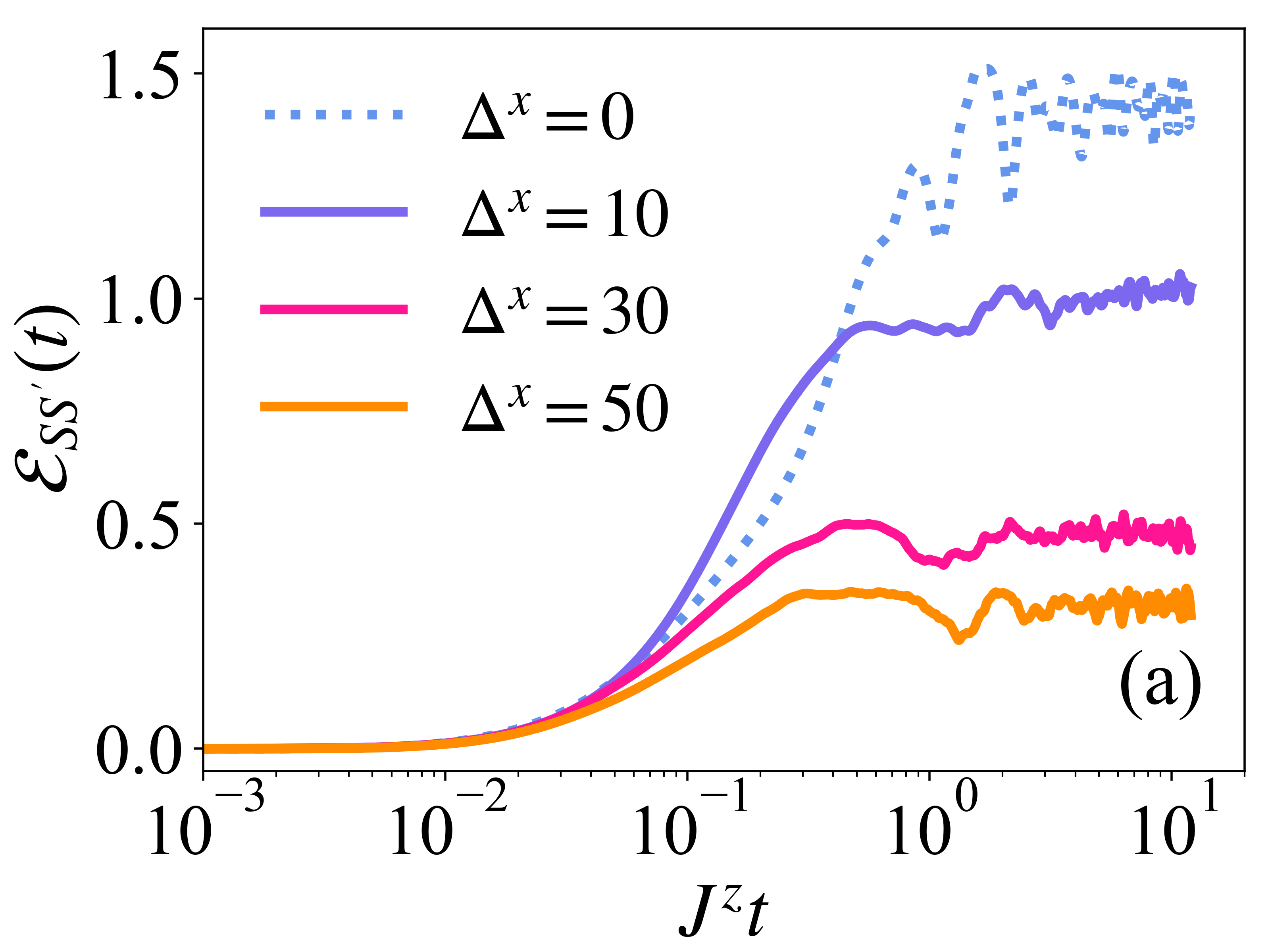}
\includegraphics[width=0.49\columnwidth]{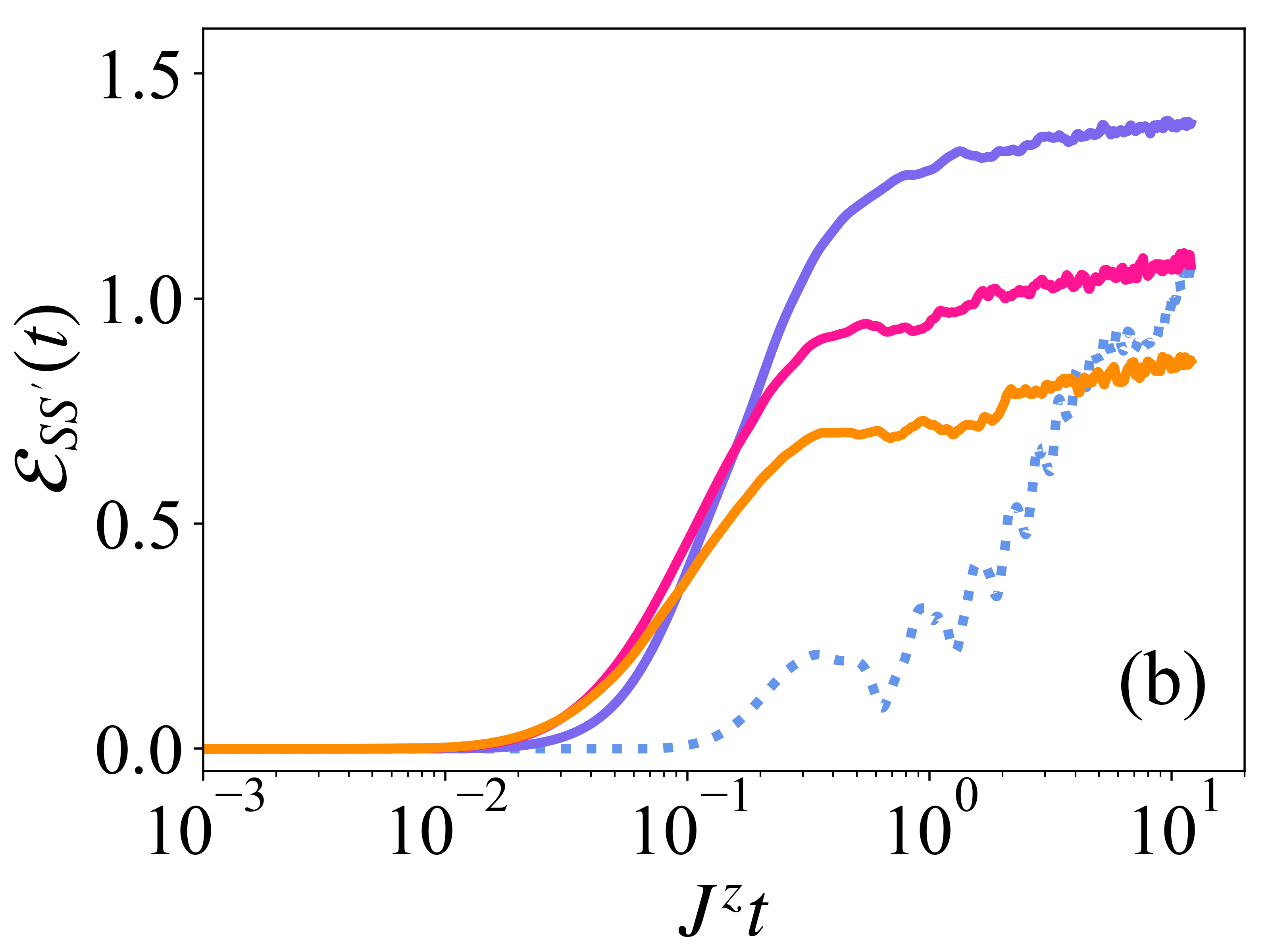}
    \caption{Logarithmic negativity for \textbf{(a)} $k=2$. \textbf{(b)} $k=7$. Plotted for $J^x=0$ and $N=8$.}
    \label{fig:fig_app_2}
\end{figure}

\end{document}